\documentclass[doublecol]{epl2} 
\title{Adiabatic quantum state transfer in tight-binding chains using periodic driving fields}
\shorttitle{Adiabatic quantum state transfer ... } %Insert here a short version of the title if it exceeds 70 characters

\author{S. Longhi \inst{1,2}}
\shortauthor{S. Longhi \etal}

\institute{                    
  \inst{1}  Dipartimento di Fisica, Politecnico di Milano, Piazza L. da Vinci 32, I-20133 Milano, Italy\\
  \inst{2}  Istituto di Fotonica e Nanotecnlogie del Consiglio Nazionale delle Ricerche, sezione di Milano, Piazza L. da Vinci 32, I-20133 Milano, Italy}

\abstract{A method for high-fidelity coherent adiabatic transport in a zig-zag tight-binding chain, based on application of two external periodic driving fields, is theoretically proposed. The method turns out to be robust against imperfections and disorder of the static lattice Hamiltonian, is tolerant to next-nearest neighborhood interactions, and enables coherent transport  in long chains without the need for a local control and timing of the trapping potential.}
\begin{document}

\maketitle

\section{Introduction}

Controlling the evolution of quantum states and the problem of quantum state transfer (QST) are of great importance in different fields of physics, with applications to e.g. quantum ratchets, mesoscopic transport and adiabatic quantum pumps, quantum entanglement  and quantum information processing \cite{R-2,R-1,R0,R1,R2,R3,R4,R5,R6,R6bis,R7,R8,R9,R10,Platero,R11,R12,R13}.  
For example, in solid-state based quantum computing it is very crucial to 
control electronic charge or spin degrees
of freedom in coupled quantum dots or spin chains, and to have a system serving
as a quantum data bus. 
So far, a wide variety of
QST schemes, based either on static or dynamically-controlled Hamiltonians, have been introduced to achieve high-fidelity transfer \cite{R3,R4,R5,R6,R7,R8,R9,R10,R11,R13,uff1,uff2,uff3,uff4,uff5,uff6}. The simplest QST methods employ static Hamiltonians and communication is achieved by simply
placing a quantum state at one end of the chain and waiting for an optimized time to let this state propagate to the other end with a high fidelity \cite{R4,R5}. Such methods, however, are 
very sensitive to imperfections of the underlying Hamiltonian and require a careful timing of the interaction. QST methods based on dynamically-controlled Hamiltonians can overcome such limitations, and have attracted increasing interest in the past recent years \cite{R7,R9,R11,uff1,uff2,uff3,uff4,uff5,uff6}.
Among different proposals of coherent QST
in time-evolving quantum systems, adiabatic methods are powerful tools
 being robust against small variations
of the Hamiltonian and the transport time. 
Coherent tunneling by adiabatic passage (CTAP), which has been independently proposed for neutral atoms
in optical traps \cite{palle1} and for electrons in quantum dot
systems \cite{palle2}, provides one of the most investigated and robust QST protocols (see, for instance, \cite{palle3,palle4,palle4bis,palle5,palle6,palle7,palle8,palle9,palle10,palle11} and references therein). Though the experimental demonstration of CTAP 
in a solid-state system is still missing,
this technique is extremely useful as a
constructive tool and is expected to play an important role in future solid-state quantum information processing.
 CTAP requires a dynamical tuning of the interaction between
adjacent quantum units following a counterintuitive scheme which is 
analog of the well-known stimulated Raman adiabatic passage (STIRAP)
protocol of quantum optics \cite{Vitanov}. Tuning of the hopping rates is generally obtained by changing
either the distance or the height of the neighboring potential
wells, i.e. it requires to reshape the trapping potential. 
 In this Letter we show that QST based on adiabatic passage can be realized in a zig-zag tight-binding chain by application of external ac control fields, without the need to modify the trapping potential. The method turns out to be robust against imperfections and disorder of the static lattice Hamiltonian, and it is tolerant to next-nearest neighborhood interactions. Our approach enables QST with high fidelity in long chains and can be useful whenever a local control and timing of the trapping potential in the various wells is unfeasible.

\begin{figure}
\onefigure[width=8.6cm]{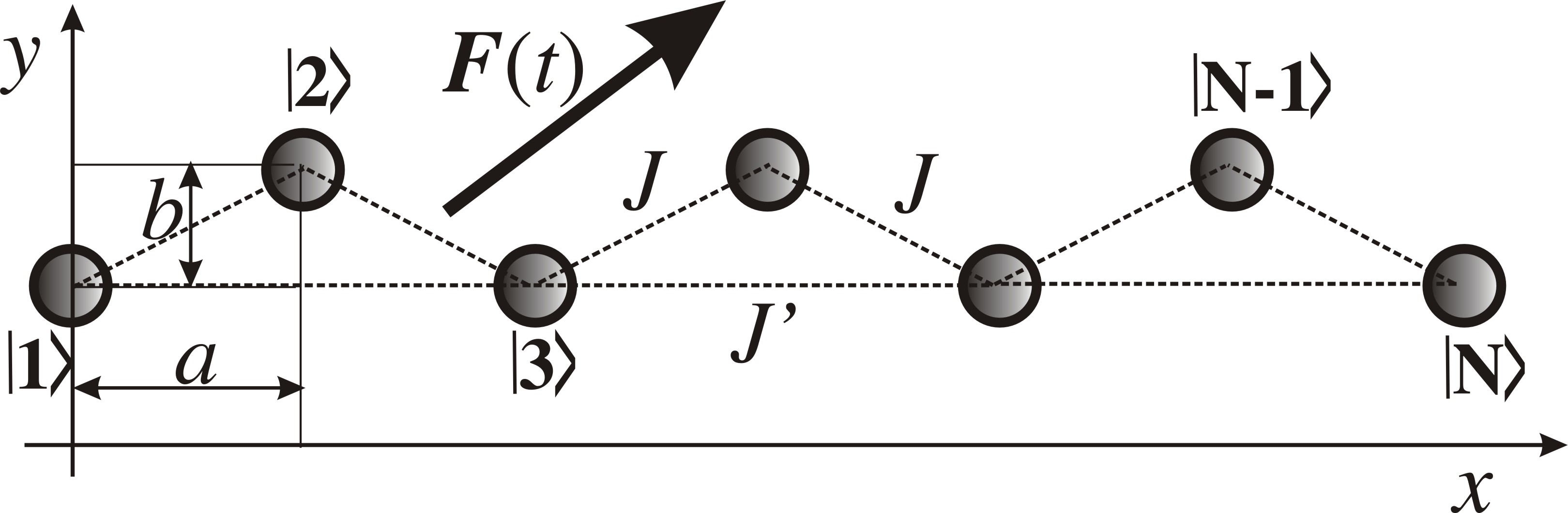}
\caption{(Color online) Schematic of a zig-zag tight-binding chain driven by an external force $\mathbf{F}(t)=(F_x,F_y)$.}
\end{figure}

\section{Coherent transport in zig-zag tight-binding chains by periodic driving fields} .\par
\par {\it The model.} Let us consider the coherent motion of a quantum particle hopping on a zig-zag chain comprising an odd number $N$ of sites, driven by an external periodic field, as shown in Fig.1. In the tight-binding approximation and assuming a coherent dynamics, the single-particle motion can be rather generally described by the tight-binding Hamiltonian (see, for instance, \cite{R7,Platero,uff6,kepalle})
\begin{eqnarray}
\hat{H} & = &  \sum_{n=1}^{N-1} J \left( | n \rangle \langle n+1|+ | n+1 \rangle \langle n| \right) \nonumber \\
& + &  \sum_{n=1}^{N-2} J'  \left( | n \rangle \langle n+2|+ | n+2 \rangle \langle n| \right) \\
& + & \sum_{n=1}^{N} \left\{ \epsilon_n+ naF_x(t)+\left[ 1+(-1)^n \right] \frac{b}{2} F_y(t) \right\} | n \rangle \langle n| \;\;\;\;\; \nonumber 
\end{eqnarray}
where $J$, $J'$ are the hopping rate between nearest and next-nearest sites, respectively, $\epsilon_n$ is the energy of site $n$, $F_x(t)$, $F_y(t)$ are the horizontal and vertical components of the external force, $a$ is the horizontal distance between adjacent sites in the chain, $b$ is the vertical displacement, and $|n \rangle$ is the Wannier state that localizes the particle at site $n$ (see Fig.1). In the following, we will assume that the potential wells in the chain are identical, and take $\epsilon_n=0$ as a reference level of energy. The forcing fields $F_x(t)$ and $F_y(t)$ are assumed to be quasi-monochromatic, with the same carrier frequency $\omega$ and slowly-varying amplitudes $A_x(t)$ and $A_y(t)$, namely
\begin{equation}
F_x(t)=A_x(t) \cos ( \omega t) \; , \;\; F_y(t)=A_y(t) \cos (\omega t).
\end{equation}
In the high-frequency limit $\omega \gg J,J'$, at leading order the role of the external forces is to effectively renormalize the hopping rates \cite{R7,Platero}. Assuming that the amplitudes $A_x$ and $A_y$ vary slowly over one oscillation cycle $2 \pi / \omega$, application of the rotating-wave approximation yields the effective Hamiltonian (see, for instance, \cite{R7,chain})
\begin{eqnarray}
\hat{H}_{eff} & = & \sum_{n=1}^{N-1} \left( \theta_n |n \rangle \langle n+1|+ \theta_n^*|n+1 \rangle \langle n| \right) \nonumber \\
& + &   \sum_{n=1}^{N-2} \left( \sigma |n \rangle \langle n+2|+ \sigma^*|n+2 \rangle \langle n| \right)
\end{eqnarray}
 with effective hopping rates
 \begin{eqnarray}
 \theta_n= \left\{ 
 \begin{array}{ll}
J \mathcal{J}_0 ( \Gamma_1) & n \; \; \rm{odd} \\
 J \mathcal{J}_0 ( \Gamma_2) & n \; \; \rm{even} \\
 \end{array}
 \right.
 \end{eqnarray}
for nearest-neighbor sites, and
\begin{equation}
\sigma=J' \mathcal{J}_0(\Gamma_3)
\end{equation}
for next-nearest-neighbor sites.
In Eqs.(4) and (5) we have set
\begin{eqnarray}
\Gamma_1(t) & = &  \frac{aA_x(t)+bA_y(t)}{\omega} \nonumber \\
\Gamma_2(t) & = &  \frac{aA_x(t)-bA_y(t)}{\omega} \\
\Gamma_3(t) & = &  \frac{2aA_x(t)}{\omega} \nonumber
\end{eqnarray}
and $\mathcal{J}_0$ is the Bessel function of first kind and zero order. Note that the effective hopping rates are slowly varying functions of time and can be controlled by appropriate shaping of the amplitudes $A_x$ and $A_y$ of the forcing fields. This enables us to effectively achieve adiabatic transport along the chain, from the first ($n=1$) to the last ($n=N$) sites, by application of external fields, without the need to deform the local trapping potential, e.g. to modify and timing distances and/or depths of the various wells which is generally required in CTAP protocols. We note that, as compared to the QST method recently proposed by Creffield in Ref. \cite{R7} using an external control field in a static bipartite lattice and based on selective destruction of tunneling, our scheme belongs to adiabatic methods, and it is thus expected to be more robust against imperfections or disorder of the static lattice Hamiltonian.\par
{\it Adiabatic state transfer protocol.}  As an example of robust adiabatic CTAP using external control fields, let us implement with our method the multilevel STIRAP protocol, first proposed for atomic systems  in Ref.\cite{Shore} and then extended to other systems, such as  linear quantum dot chains \cite{palle4bis} and optical waveguide arrays \cite{LonghiPLA}. The basic principle of adiabatic quantum transport in the multilevel STIRAP protocol is based on the existence of a dark state, with energy $E=0$, for the Hamiltonian (3) when next-nearest-neighbor tunneling is negligible. In this case, after setting in Eq.(3) $\sigma=0$ and assuming $\theta_n=\Omega_1$ for $n$ odd and $\theta_n=\Omega_2$ for $n$ even, it can be readily shown that the Hamiltonian $\hat{H}_{eff}$ admits of an instantaneous eigenstate $| \psi_0 \rangle$ with energy $E=0$, called dark state and given by \cite{Shore}
\begin{equation}
\langle n | \psi_0 \rangle= \left\{
\begin{array}{ll}
0 & n \; \; {\rm even} \\
\left[ - \frac{\Omega_1}{\Omega_2} \right]^{(n-1)/2}  \times \\
\sqrt{\frac{1-[\Omega_1/ \Omega_2]^2}{1-[\Omega_1/ \Omega_2]^{N+1}}} 
& n \; \; {\rm odd}
\end{array}
\right.
\end{equation}
Note that, for $\Omega_1 / \Omega_2 \rightarrow 0$ the dark state $| \psi_0 \rangle$ localizes the particle at the left boundary site $n=1$, i.e. $\langle n | \psi_0 \rangle \simeq \delta_{n,1}$, whereas for $\Omega_1 / \Omega_2 \rightarrow \infty$ the dark state corresponds to the particle being localized at the right boundary of the chain, i.e.  $\langle n | \psi_0 \rangle \simeq \pm \delta_{n,N}$. Perfect particle transfer from the site $n=1$ to the site $n=N$ can be thus obtained by adiabatic change of the tunneling amplitudes $\Omega_{1,2}$ such that $\Omega_1 / \Omega_2 \rightarrow 0$ at initial time and  $\Omega_2 / \Omega_1 \rightarrow 0$ at final time. This can be accomplished by assuming
a so-called counter-intuitive sequence \cite{Shore} $\Omega_1(t)=\Omega(t+\delta/2)$ and $\Omega_2(t)=\Omega(t-\delta/2)$, where $\Omega(t)$ is a bell-shaped and slowly-varying function of time with $\Omega(t) \rightarrow 0$ as $t \rightarrow \pm \infty$ and $\delta>0$ is a time delay. The conditions that ensure adiabatic evolution of the system in the dark state $| \psi_0 \rangle$ are discussed in Ref.\cite{Shore}; in particular for a finite transit time $\tau_t=2T$ it turns out that the optimized time delay $\delta$ of the counterintuitive pulse sequence is of the same order than the duration $\tau$ of the pulse $\Omega(t)$.  Note that the state transfer can be reversed by changing the sign of $\delta$.
To realize perfect adiabatic state transfer in our zig-zag chain,  let us assume that direct tunneling to next-nearest-neighbors is negligible over the transfer time $\tau_t=2T$, i.e. $2 J' T \ll 1$, and let us tailor the force amplitudes $A_x(t)$ and $A_y(t)$ such that  $\theta_n(t)=\Omega(t-\delta/2)$ for $n$ odd, and $\theta_n(t)=\Omega(t+\delta/2)$ for $n$ even, where $\Omega(t)$ is a bell-shaped function. According to Eqs.(4) and (6), this can be achieved by taking, for example, Gaussian-like shapes for the amplitudes $A_x(t)$ and $A_y(t)$, namely 

\begin{figure}
\onefigure[width=8.8cm]{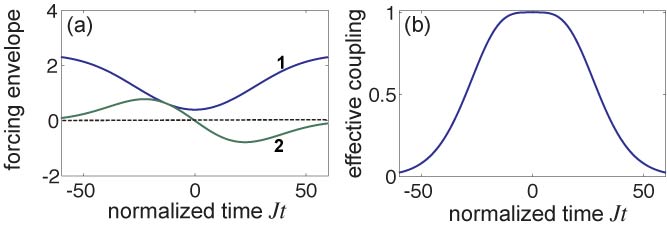}
\caption{(Color online) (a) Typical behaviors of the normalized forcing envelopes $aA_x(t)/ \omega$ (curve 1)  and $bA_y(t) / \omega$ (curve 2) that realize adiabatic multilevel CTAP [see Eqs.(8) and (9)]. Parameter values  are $TJ=60$, $\tau/T=0.5$, $\delta / \tau=0.85$ and $\omega /J=10$. (b) Corresponding behavior of the effective hopping rate $\Omega(t)$, in units of $J$ [Eq.(10)].}
\end{figure}

\begin{figure}
\onefigure[width=7cm]{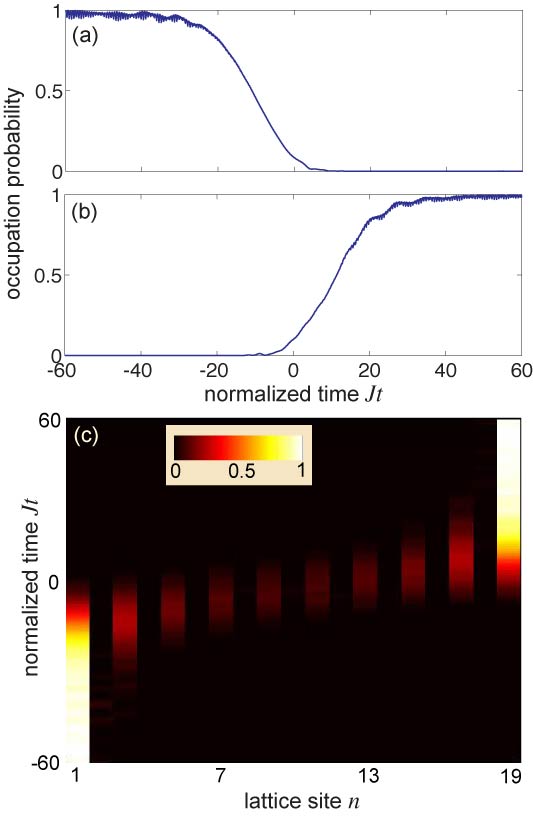}
\caption{(Color online) CTAP realized in a chain of $N=19$ sites by the external forcing of Fig.2 and for $J'=0$. (a), (b) show the evolution of the occupation probabilities $|\langle 1 | \psi(t) \rangle|^2$ and $| \langle N | \psi(t) \rangle|^2$ of sites $n=1$ and $n=N$, respectively. In (c) the evolution of occupation probabilities $| \langle n | \psi(t) \rangle|^2$ of the various sites of the chain are shown in a pseudocolor map.}
\end{figure}

\begin{eqnarray}
A_x(t) & = & \frac{2.405 \omega}{2a} \left\{ 2- \exp[-(t-\delta/2)^2 / \tau^2] \right. \nonumber \\
& - & \left.  \exp[-(t+\delta/2)^2 / \tau^2] \right\} \\
A_y(t) & = & \frac{2.405 \omega}{2b} \left\{ \exp[-(t+\delta/2)^2 / \tau^2] \right. \nonumber \\
& - & \left. \exp[-(t-\delta/2)^2 / \tau^2] \right\} 
\end{eqnarray}
 corresponding to 
 \begin{equation}
 \Omega(t)=J \mathcal{J}_0 \left( 2.405 \left\{ 1-\exp(-t^2/\tau^2) \right\} \right),
 \end{equation}
with $\delta <  \sim \tau $ and $\exp(-T^2/\tau^2) \ll 1$. A typical behavior of $A_x(t)$, $A_y(t)$ and $\Omega(t)$ in the interval $(-T,T)$ is shown in Fig.2. Note that at $t \rightarrow \pm \infty$ one has $aA_x/ \omega \rightarrow 2.405$ (the first root of the $\mathcal{J}_0$ Bessel function) and $bA_y/ \omega \rightarrow 0$, corresponding to a regime of frozen dynamics (because of coherent destruction of tunneling induced by the $A_x$ force component). This means that the external field $A_x$ should be switched on before the system is prepared with the particle in the initial site $|1 \rangle$, and should not turned off when the transfer to the site $|N \rangle$ has been achieved. 
It should be also noted that the adiabatic transport method using ac control fields proposed in this work differs from  parametric quantum pumping methods in mesoscopic open systems \cite{R-1,R0} on several and important instances. In quantum pumps, 
average current between two reservoirs
that are kept at the same bias is obtained by {\it slowly}  and {\it cyclically} varying 
at least two parameters of the system. This mechanism was originally described by
Thouless \cite{R-2} for isolated (or otherwise gapped)
systems at zero temperature and is based on a periodic  
and {\it slow} variation in time of the scattering properties of the pump region. 
Generally this is obtained by a deformation of the confining potential that is slow compared
with the relevant energy relaxation times. 
A net current is generated by varying
in a {\it cyclic} fashion and {\it out of phase} two independent parameters that control the confining potential, i.e. the system Hamiltonian \cite{R-2,R-1,R0}.  In our scheme, the quantum system is closed, quantum transfer 
does not require any deformation of the confining potential, the periodic modulation introduced by the two external fields is
{\it fast} ( as compared with the typical frequency of the system given by the hopping rate), and the applied fields are {\it in phase} [rather than out of phase; see Eq.(2)]. 
Such main differences stem from the fact that the mechanism of adiabatic quantum transport in our case is not based on a cyclic and slow evolution of the system Hamiltonian, rather it realizes an adiabatic and {\it non-cyclic} evolution of the effective system Hamiltonian (3) in a dark state \cite{Shore}. The in-phase high-frequency fields provide the appropriate tailoring of the effective hopping rates entering in Eq.(3). Basically, at $ t \rightarrow \pm \infty$ the oscillating force $F_y(t)$ is switched off whereas the oscillating force $F_x(t)$ is switched on with an amplitude that realizes coherent destruction of tunneling between adjacent sites: in this regime the dynamics is basically frozen, provided that next-nearest-neighbor hopping is negligible. To realize adiabatic transfer, the force $F_x(t)$ is diminished whereas the force $F_y(t)$ is switched on, following the characteristic even/odd profiles depicted in Fig.2(a). The combined forcing in the horizontal and vertical directions yields effective asymmetric tunneling rates at alternating sites, mimicking the counterintuitive scheme of multilevel STIRAP \cite{Shore}. \par
{\it Numerical simulations.} We checked the predictions of the theoretical analysis and the fidelity of the adiabatic transfer method by direct numerical simulations of the single-particle Schr\"{o}dinger equation $ i \partial_t | \psi(t) \rangle  = \hat{H}(t) | \psi(t) \rangle$ using the original Hamiltonian (1) with $J'=0$, i.e. neglecting next-nearest-neighbor hopping in the chain. At initial time $t=-T$ the system is prepared in the state $|1 \rangle$, i.e. $|\psi(-T) \rangle=|1 \rangle$, and we wish to transfer the excitation to the site $n=N$ after the transfer time $\tau_t=2T$ with high fidelity, i.e. $|\psi(T) \rangle =|N \rangle$. 
Figure 2(a) shows a typical behavior of the driving forces used in the numerical simulations, which have been assumed to follow the profiles defined by Eqs.(8) and (9). Parameter values used in the simulations are $\omega / J=10$, $J T=60$, $\tau/T=0.5$  and $\delta/ \tau=0.85$. High-fidelity adiabatic passage from the site $n=1$ to the site $n=N=19$ in a chain comprising $N=19$ sites is shown in Fig.3. Figures 3(a) and 3(b) show the detailed temporal evolution of the occupation probabilities of boundary sites $n=1$ and $n=N$, whereas Fig.3(c) shows in a pseudocolor map  the evolution of occupation probabilities of all sites. Note that, according to the adiabatic analysis, excitation of the sites with even numbers is negligible, indicating that the system evolves following the zero-energy  adiabatic state (7). The final occupation probability of site $N$ is $>99 \%$, indicating a high fidelity of the transfer process. Note that the transfer time $\tau_t=2 T=120/ J$ is about $4.2$ longer than the fastest time $\tau_{min} \simeq  \pi (N-1)/(2J)$ that one could achieve using a non-adiabatic method by sequentially
pulsing the tunneling rates between adjacent sites. Nevertheless, the adiabatic nature of the transfer process makes it rather robust against imperfections of the static lattice Hamiltonian, as shown in the next section.

\begin{figure}
\onefigure[width=7cm]{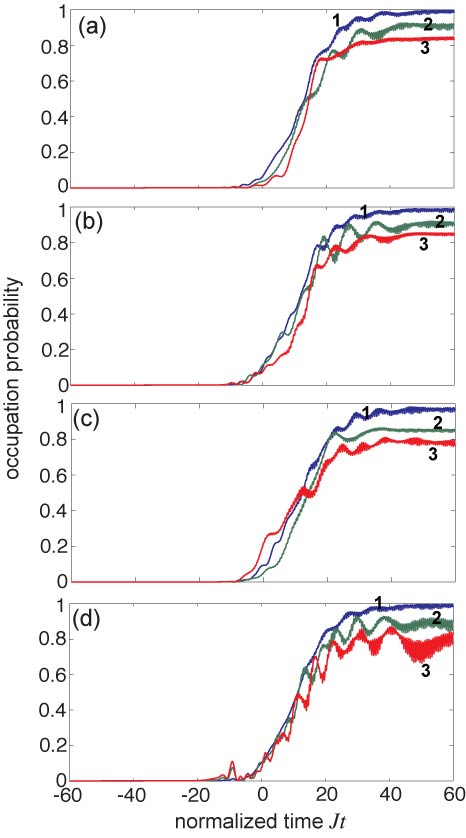}
\caption{(Color online) Effects of lattice disorder and non-nearest-neighbor hopping on adiabatic QST. The panels show the evolution of the occupation probability $|\langle N | \psi(t) \rangle|^2$ of site $N=19$ in the same zig-zag chain of Fig.3 and for the same parameter values (a) in presence of disorder of hopping rates (curve 1: $\Delta=0.1$, curve 2: $\Delta=0.2$, curve 3: $\Delta=0.3$), (b) in presence of site energy disorder (curve 1: $\Delta=0.1$, curve 2: $\Delta=0.2$, curve 3: $\Delta=0.3$),  (c) in presence of simultaneous disorder of hopping rates and site energies (curve 1: $\Delta=0.1$, curve 2: $\Delta=0.2$, curve 3: $\Delta=0.3$), and (d) in presence of a non-vanishing next-nearest-neighbor hopping $J'$ (curve 1: $J'/J=0.05$, curve 2: $J'/J=0.1$, curve 3: $J'/J=0.15$).}
\end{figure}

\section{Effects of lattice disorder and next-nearest-neighbor hopping} 
In the previous section we have shown, both analytically and numerically, that two external driving fields in a zig-zag tight-binding chain can realize long-distance CTAP with high fidelity. 
Since the transport mechanism is based on an adiabatic
process, it is expected to be robust against inhomogeneities or
disorder of the static lattice. We have checked the robustness of our proposed scheme by direct
numerical simulations of the single-particle Schr\"{o}dinger equation for the ac-driven zig-zag chain, described by the Hamiltonian (1) with $J'=0$, in the
presence of disorder for either the hopping rate $J$ and the on-site
energies $\epsilon_n$. In a first set of simulations, we considered disorder in the hopping rate and
assumed $ J \rightarrow J(1+\delta _n )$ in Eq.(1), where $\delta_n$ is a random variable
with zero mean and uniform distribution in the range $(- \Delta, \Delta)$,
with $\Delta< 1$. Figure 4(a) shows, as an example, the numerically-computed evolution of the 
occupation probability of site $N$, for the same zig-zag chain and driving conditions used in Fig.3 and for three 
realizations corresponding to increasing values of the disorder $\Delta$. The figure clearly shows that the adiabatic transfer method is rather robust against disorder of the hopping rates, with a fidelity that remains larger than $ \sim 90 \%$ for a disorder of $\sim 20 \%$. In a second set of simulations, we studied the impact of disorder of site energies $\epsilon_n$ on the adiabatic transfer. Different site energies $\epsilon_n$ arise, in practice, owing to differences in the confining potential of the various wells in the chain. The disorder is simulated by assuming $\epsilon_n=J \delta _n$, where $\delta_n$ is a random variable
with zero mean and uniform distribution in the range $(- \Delta,\Delta)$.  As an example, Fig.4(b) shows typical results of the adiabatic transfer as obtained for increasing values of the disorder $\Delta$. Note that a fidelity larger than $ 90 \%$ is observed for a disorder of the site energies smaller than $\sim 20\% $, indicating that the QST method is rather tolerant to moderate inhomogeneities of the site energies as well. Finally, Fig.4(c) shows the results of QST in the presence of simultaneous disorder in both site energies and hopping rates.  A high fidelity ($> 95 \%$) is observed in this case for a small to moderate disorder [see curve 1 in Fig.4(c)].\par
We also investigated the impact of next-nearest-neighbor hopping on the adiabatic transfer process. Like for other adiabatic and non-adiabatic QST methods, the presence of 
next-nearest-neighbor hopping is detrimental and generally it should be avoided by either reducing the transfer time $2T$ or minimizing the direct tunneling of distant potential wells. In our zig-zag geometry,  minimization of the ratio $J'/J$ requires to operate in a geometrical setting with $b \ll a$. Figure 4(d) shows the impact of next-nearest-hopping on the QST process for the same parameter values of Fig.3 and for increasing values of $J'/J$. The results clearly indicate the detrimental effect of next-nearest-neighbor hopping, however a fidelity larger than $ \sim 90 \%$ is still observed provided that $J'/J  < 0.1$.
\section{Experimental implementation} Let us finally briefly discuss possible physical realizations of the adiabatic quantum state transfer method discussed in the previous sections. The Hamiltonian (1) can be implemented in different physical systems, for example using dilute ultracold atoms or trapped ions in zig-zag optical lattices \cite{zigzag1,zigzag2,zigzag3}, arrays of superconducting flux quantum bits with programmable spin-spin
couplings \cite{super}, charge transport in zig-zag quantum dot chains \cite{zigzag4,zigzag5}, and photonic transport in zig-zag evanescently-coupled optical waveguide arrays with periodically-bent axis \cite{Longhi1,Longhi2}. The transit time $\tau_t=2 T$ should be smaller than the decoherence time of the system, yet long enough to ensure adiabatic evolution of the system in its dark state. Let us consider, for example, an implementation of the Hamiltonian (1) based on dilute ultracold atoms in zig-zag optical lattices \cite{zigzag1,zigzag2,zigzag3}.  In such a system, the external ac fields can be realized by periodically shaking the optical lattice in both $x$ and $y $ directions \cite{Ari}. A typical
value of the hopping rate is $J \simeq 10^4 \; {\rm rad/s}$ \cite{Folling}. Correspondingly, the adiabatic transfer shown in Figs.2 and 3 corresponds, in physical units, to a modulation frequency $\omega/ ( 2\pi) \simeq 15.9$ kHz of the ac force, a pulse duration $ \tau \simeq 1.5 \;$ms and a pulse delay $\delta \simeq 1.275$ ms.
The transit time $\tau_t=2T$ is $\tau_t \simeq 12$ ms, which is much smaller
than the decoherence time (hundreds of ms) determined by the
typical lifetime due to inelastic scattering of lattice photons \cite{Folling}.

\section{Conclusions} In this work we have introduced a method for high-fidelity coherent adiabatic transport in a static zig-zag tight-binding chain, based on application of two external periodic driving fields. The method turns out to be robust against imperfections and disorder of the static lattice Hamiltonian, is tolerant to next-nearest neighborhood interactions, and enables coherent transport  in long chains without the need for a local control and timing of the trapping potential wells. Our setup provides a novel route to implement CTAP protocols using oscillating driving fields and  can be of interest whenever control and timing of the trapping potential in the various wells in unfeasible.

%\acknowledgments

\end{document}